# DESIGN STUDY OF A BETA=0.09 HIGH CURRENT SUPERCONDUCTING HALF WAVE RESONATOR *

H.T.X Zhong, F Zhu[#], P.L Fan, S.W Quan, K.X Liu

SKLNST & IHIP, School of Physics, Peking University, Beijing 100871, China

## Abstract

There's presently a growing demand for high current proton and deuteron linear accelerators based on superconducting technology to better support various fields of science. A β=0.09 162.5 MHz high current superconducting half wave resonator (HWR) has been designed at Peking University to accelerate 100 mA proton beam or 50 mA deuteron beam after the RFQ accelerating structure. The detailed electromagnetic design, multipacting simulation, mechanical analysis of the cavity will be given in this paper.

## INTRODUCTION

There's presently a growing demand for high current proton and deuteron linear accelerators to better support various fields of science. More and more projects based on such machines have emerged and been proposed, such as CADS of 10mA proton beam, Beijing Isotope-Separation-On-Line neutron beam facility (BISOL) of 10mA deuteron beam [1, 2], and IFMIF of 125mA deuteron beam for one linear accelerator [3]. After RFQ, the beams will be accelerated by a RF superconducting (SRF) linear accelerator to get high current and to the desired energy. Comparing to superconducting quarter wave resonator (QWR), half wave resonator (HWR) has symmetrical fields thus no dipole steering to the beam and is better for high current ion beam acceleration. HWR also has better mechanical properties. A β=0.09 162.5 MHz HWR cavity has been designed to accelerate 100mA proton beam or 50 mA deuteron beam after RFQ. We will present the details of design work of the β=0.09 HWR cavity in this paper.

## ELECTROMAGNETIC DESIGN

The electromagnetic design optimization is mainly the minimization of the peak surface fields over the gradient $E_{pk}/E_{acc}$ and $B_{pk}/E_{acc}$, and maximization of the shunt impedance r/Q and geometry factor G. Compared to the cylindrical or squeezed HWR cavity [4], the taper type HWR cavity which has conical inner and outer conductors has much higher r/Q and lower surface fields, especially magnetic surface field when the gradient is fixed [5]. The electromagnetic design was done by CST code [6]. Fig. 1 shows the simulation model of the HWR cavity.

In order to accelerate 100 mA proton beams, the diameter of the beam pipe is an important parameter. Larger diameter leads to worse RF properties of the cavity. But when the diameter of the beam port is small, beam loss for high current beams may be serious. Fig. 2 shows the transverse component field of two HWR cavities with different diameters of the beam pipes. When the beam is off axis, the transverse component filed of the HWR with smaller beam pipe is stronger than that at the larger beam pipe case when the gradients are the same. After simulation, we decide the diameter of the beam pipe to be 40mm. Larger diameter of the outer conductor is nice for lower $B_{pk}/E_{acc}$, but it is not economic. 260mm is chosen for balance. The iris-to-iris length is 2βλ/3.

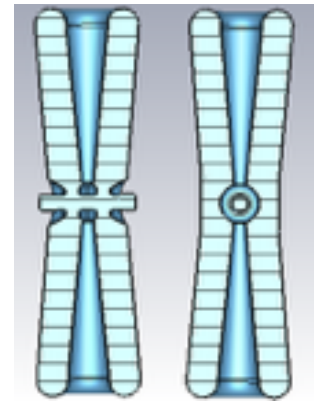

Figure 1: Model for the β=0.09 162.5 MHz HWR in CST code

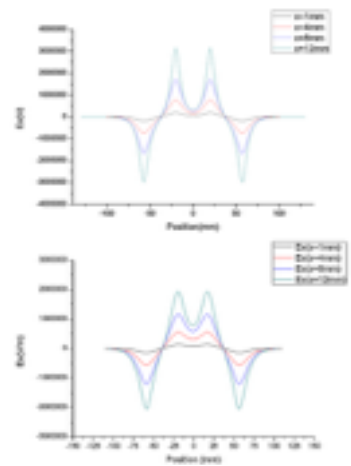

Figure 2: Transverse component fields of HWR with a beam pipe diameter of 30mm (top) and HWR with a diameter of 40mm (bottom)

Compared to the race-track shaped center conductor, the ring-shaped "Donut" center conductor has much lower peak magnetic field and thus higher accelerating gradient and much higher shunt impedance meaning same energy gain less power [7]. With this "Donut" shape, there is better symmetric field in radial direction along the beam pipe and can eliminate the quadrupole effect to the beam. Table 1 gives the main geometry and RF parameters of the HWR cavity.

___
*Work supported by National Basic Research Project (No. 2014CB845504)
[#]zhufeng7726@pku.edu.cn

Table 1: RF and geometry parameters of the HWR cavity

| Parameter | Value |
|---|---|
| Frequency/MHz | 162.5 |
| Optimal β | 0.09 |
| Cavity diameter /mm | 260 |
| Beam aperture /mm | 40 |
| Cavity height /mm | 990 |
| $L_{cav}=\beta\lambda$ /mm | 166 |
| R/Q /Ω | 255 |
| Geometry factor /Ω | 39 |
| Bpk/Eacc /(mT/(MV/m)) | 6.4 |
| Epk/Eacc | 5.3 |

## MULTIPACTING SIMULATION

One of the main limitations for low β SRF cavity is multipacting (MP). We use CST particle tracking mode to do the MP simulation. The simulation result gives that MP in the HWR cavity mainly locates at the dome of the short plate as shown in Fig. 3. It is two-point first order multipacting, which agrees with the previous studies on the similar HWRs [8]. Fig. 4 shows the simulation result for the cavity HWR1 with round short plate. The radius of the round short plate is 39 mm. When the gradient is in the range of 3~8MV/m, the MP possibility of HWR1 is high, therefore we need to eliminate the potential MP barriers.

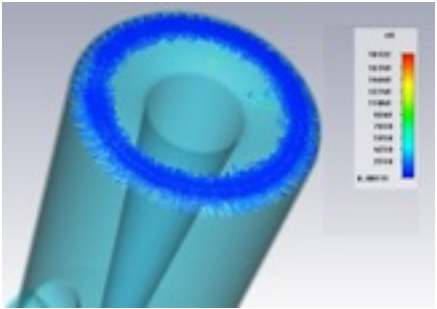

Figure 3: Location of MP electrons near the short plate.

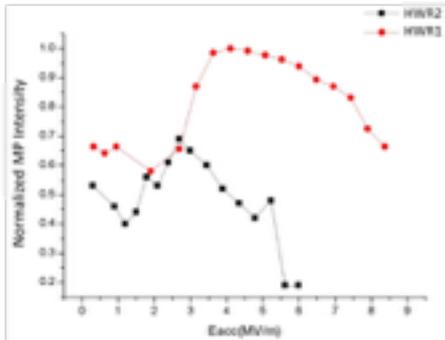

Figure 4: Normalized MP Intensity v.s. cavity gradient for HWR1 and HWR2.

Once the electrons moving around with different geometry, it becomes more difficult to maintain in the resonant condition and MP can be suppressed. Short height of the dome helps to suppress MP [9]. We changed the blending radii of the short plate with the inner and outer conductors and make the short plate flatter. The shape of the short plate is seen in Fig. 5. R1 is the blending radius with the inner conductor and r2 with the outer conductor. Fig. 6 shows the simulation results. From the curves, we can see that smaller r1 and larger r2 has better effect of suppressing MP. When r1 = 5 mm and r2 =35 mm, The MP intensity is quite low when the gradient is higher than 4 MV/m, which is safe for the cavity operation. Fig. 4 gives the MP intensity comparison between HWR1 with round short plate and HWR2 which has flat short plate with r1 = 5 mm and r2 =35 mm. Calculation from CST shows that the RF parameters are similar for the two HWR cavities.

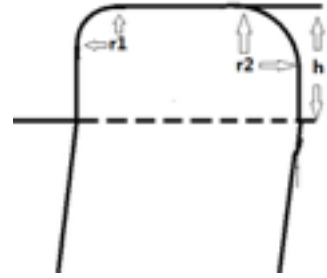

Figure 5: Shape of the short plate with different blending radii with the conductors.

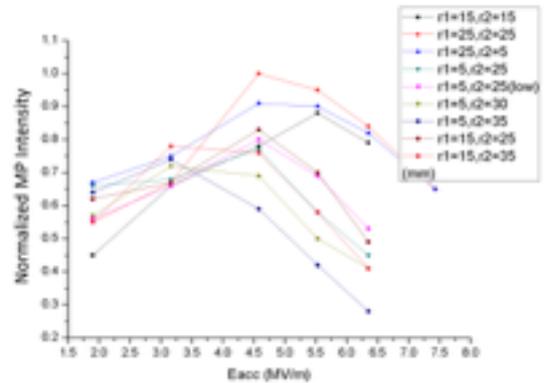

Figure 6: MP Intensity v.s. gradient for different cavities with short plates of various blending radii.

## MECHANICAL ANALYSIS

Another important feature is the mechanical property of the HWR cavity. The main elements of mechanical instability for SRF cavity are microphonics and Lorentz force detuning. Pressure fluctuation is one main source for microphonics. The frequency shift caused by helium bath pressure and Lorentz force is analyzed with ANSYS [10, 11]. The pressure sensitivity coefficient, df /dp, is used to characterize the influence of the helium pressure fluctuation on the detuning of the cavity. The Lorentz force detuning coefficient $K_L=df/E_{acc}^2$ is used to describe the effect of the Lorentz force on the detuning of the cavity. Fig. 7 shows cavity deformation results by pressure at different places for HWR1 and HWR2. Table

2 gives the mechanical parameters of these two HWR cavities.

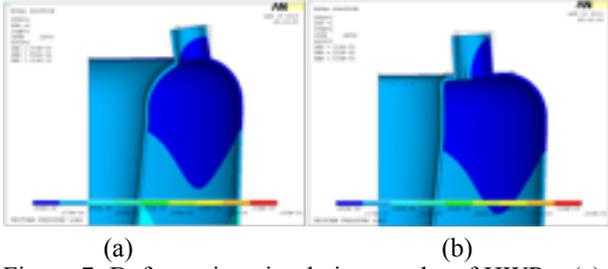

(a)          (b)

Figure 7: Deformation simulation results of HWRs: (a) at the round short plat of HWR1, (b) at the flat short plate of HWR2.

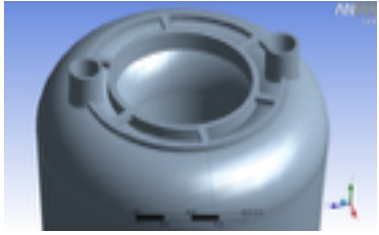

Figure 8: The stiffening rings at the short plate.

Table 2: Mechanical properties of HWR1 and HWR2

| Cavity with different boundary condition | df/dp (Hz/mbar) | $K_L$ (Hz/(MV/m)$^2$) |
|---|---|---|
| HWR1 @ beam ports free | -36.2 | -5.9 |
| HWR1 @ beam ports fixed | 2.8 | -0.4 |
| HWR1 with stiffening rings @ beam ports fixed | -0.19 | -0.3 |
| HWR2 with stiffening rings @ beam ports fixed | 0.01 | -0.3 |
| HWR2 @ beam ports free | -36.0 | -5.98 |
| HWR2 @ beam ports fixed | 3.0 | -0.41 |

When the beam ports are free, the main deformation caused by pressure locates at the electric filed area. The electric filed area deformation drops the frequency and df/dP is negative. The deformation near the magnetic field area increases the frequency. When the beam ports fixed, the deformation near the magnetic field area dominates and df/dP is positive. By adding stiffening rings at the short plate, df/dP can change to near 0, which is very helpful for cavity operation. Fig. 8 shows the shape of the stiffening rings at the short plate area. Compared to HWR1, HWR2 has slightly larger df/dP when the beam ports are fixed and the main deformation is near the short plate. From Table 2, we can see that the mechanical properties have no big difference between HWR1 and HWR2.

The maximum deformation caused by Lorentz force is also near the electric field area. Lorentz force detuning coefficient $K_L$ is ~ -6 Hz/(MV/m)$^2$ when the beam ports are free for both HWR1 and HWR2. But when the beam ports are fixed, |$K_L$| is smaller than 1.

## EFFECT OF RINSE PORTS

Rinse ports at the short plates are necessary for cavity post-treatment. Large aperture of the rinse ports is good for effective cleaning. But the distance between the inner and outer conductor at the short plate is only 78mm for the β=0.09 HWR cavity and even smaller in the middle part of the taper type cavity. The surface magnetic field can be affected strongly by adding the rinse ports. Fig. 9 (a) shows $B_{pk}/E_{acc}$ changing with the blending radius of the rinse ports. When the blending radius of the rinse ports is 3mm which is easy for manufacture, $B_{pk}/E_{acc}$ of the cavity is high as 7.4 mT/(MV/m). When the blending radius is larger than 6mm, $B_{pk}/E_{acc}$ reduces to the level without rinse ports. We decide the final blending radius of the rinse ports to be 6mm. Fig. 9 (b) shows $B_{pk}/E_{acc}$ changing with the position of the rinse port in radial direction. We can see $B_{pk}/E_{acc}$ decreases as the rinse ports farther from the inner conductor. The $B_{pk}/E_{acc}$ increase to 6.42 mT/(MV/m) from 6.22 mT/(MV/m) for the final design with the rinse ports. The diameter of the rinse ports is chosen to be 30mm. The position of the rinse port is 85mm in radial direction.

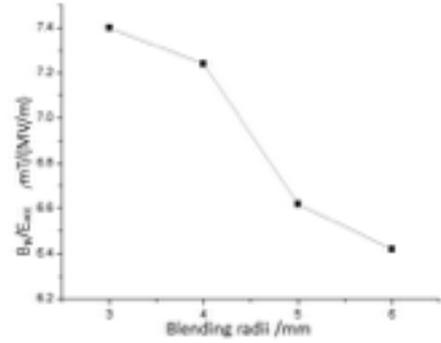

(a)

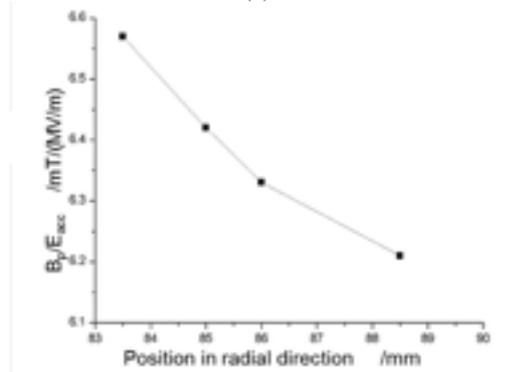

(b)

Figure 9: (a) $B_{pk}/E_{acc}$ v.s. the blending radius of the rinse ports, (b) $B_{pk}/E_{acc}$ v.s. the position of the rinse port in radial direction.

MP simulation for the HWR with the rinse ports is also analyzed. Fig. 10 shows the simulation result. We can see that there is not much difference of the MP for HWR2 with or without the rinse ports.

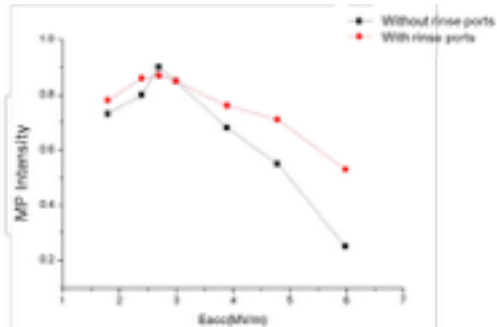

Figure 10: MP intensity for HWR2 with and without rinse ports.

## CONCLUSION

A β=0.09 162.5MHz HWR cavity has been designed at Peking University for the high current ion accelerator. We have finished the optimization of the electromagnetic parameters, MP simulation and mechanical analysis of the HWR cavity. We compared two HWR cavities with different short plates. HWR2 with flat short plate and asymmetric blending radius with the inner and outer conductors is preferred because it can suppress MP better. Mechanical calculation gives that the taper type HWR cavity has very low df/dP and Lorentz force detuning coefficient when the beam ports are fixed. The fabrication of HWR2 is under construction. The beam dynamic simulation of this high current HWR cavity will be done soon.